\documentstyle[12pt]{article}
\begin{document}
\vspace{20mm}
\begin{center}
{\large {\bf The Hawaii K-Band Galaxy Survey. II. Bright K-band Imaging}}\\
\vspace{10mm}
J.-S. Huang, L. L. Cowie, 
J. P. Gardner\footnote{Current address: Department
of Physics, University of Durham, South Road, Durham, DH1 3LE, UK.},
E. M. Hu, A. Songaila, and R. J. Wainscoat\\
Institute for Astronomy, University of Hawaii, 2680 Woodlawn Drive, Honolulu,
HI 96822\\
\vspace{10mm}
Recieved\underline {~~~~~~~~~~~~~~~~~~~~~~~~~~~~~~~~};~~~~~~accepted\underline {
~~~~~~~~~~~~~~~~~~~~~~~~~~~~~~~~}\\
\vspace{10mm}
Accpeted by {\sl Astrophysical Journal}\\
\end{center}

\newpage
\begin{abstract}

  We present the results of a wide-field K-selected galaxy survey with 
complementary optical
I- and B-band imaging in six fields with a total coverage
of 9.8 square degrees. 
The observations
were carried out on the UH 0.6m and the UH 2.2m telescopes.
The purpose of this survey is to study the properties of the local
galaxies and
explore the evolution of K-selected galaxies at low redshifts.
Star-galaxy discrimination is performed using both galaxy color properties
and object morphologies, and 6264 galaxies are found.
This survey establishes the bright-end K-band galaxy number
counts in the magnitude range $13<K<16$ with high precision. 
We find that our bright-end counts have a significantly steeper
slope than the prediction of a no-evolution model, which
cannot be accounted for by known observational or theoretical error. 
We also argue against the likelihood of sufficient evolution at
such low redshifts to account for this effect,
we describe an alternative picture in which
there is a local deficiency of galaxies by a factor of 2
on scale sizes of around 300$h^{-1}$ Mpc. Taken at face value, 
this would imply that local
measurements of $\Omega_0$ underestimate the true value of the cosmological
mass density by this factor and that local measurements of $H_0$ could be
high by as much as 33\%.\\
\begin{flushleft}
{\bf Subject headings:} galaxies: evolution $-$ galaxies: photometry $-$ infrared: galaxies\\
\end{flushleft}
\end{abstract}
\vspace{5mm}
\begin{center}
{\bf 1. Introduction}
\end{center}
\vspace{5mm}

 Near-infrared galaxy surveys (Mobasher {\sl et al.}  1986, 
Gardner {\sl et al.}  1993, Glazebrook
{\sl et al.}  1994, Cowie {\sl et al.}  1994, McLeod {\sl et al.}  1995,
Djorgovski {\sl et al.}  1995) have provided extensive information
on galaxy evolution. As noted by these studies, a near-infrared selected 
galaxy sample
may provide more advantages in studying galaxy evolution and cosmological
geometry than an optically selected sample. Its insensitivity to 
transient burst of star formation and 
its small K-corrections even at large redshift
simplify our modeling of the galaxy number count-magnitude relation, 
and can lead
to a direct determination of galaxy evolution. The near-infrared 
K-correction is also insensitive to spectral type, thus a near-infrared 
selected sample is unbiased with respect to the mix of spectral type.
This is very valuable in studying large-scale structure.\\

  Because of the small format of the first generation IR arrays,  
the early K-band surveys ( Gardner {\sl et al.}  1993, Glazebrook 
{\sl et al.}  1994, Cowie {\sl et al.}  1994) were mainly deep surveys
over relatively small areas.
After the Keck telescope
became available, it became easier to conduct 
such pencil beam deep K-band surveys
(Djorgovski {\sl et al.}  1995). Unlike 
the B-band number counts, which show a large excess
over the no-evolution prediction for the faint-end counts, 
the K-band number
counts established from these surveys do not show such a large excess. 
However, in order to draw conclusions on the nature of
galaxy evolution in these deep samples, it is essential 
to have extensive knowledge of the
local K-band luminosity function and the color distribution of a
local K-selected sample and there are few
such samples available. The primary sample used as
a local reference was one obtained
by Mobasher {\sl et al.} (1986). Their complete K-band sample to K=12 was
extracted from the B-selected sample. Later a K-band local
luminosity function was constructed  using this sample
(Mobasher {\sl et al.}  1993). \\

   The recent development of 
large format IR-arrays such as the NICMOS (256$\times$256) and HAWAII
(1024$\times$1024) HgCdTe devices, has now made it possible to conduct a large
field near-infrared survey. The Hawaii wide-field K-band survey aims to
study the local galaxy properties in the near-infrared.
With about 10 square degrees coverage and a limiting 
magnitude of K=16, the sample contains
6264 galaxies.  This sample will allow us to determine the local galaxy
density and the morphological mix  of the K-selected local galaxy sample.
The current availability of a multi-fibre spectrograph (Taylor 1994)
also makes it possible to conduct redshift measurements effectively for
our sample. Hence, using this sample we will eventually be able to construct 
a more precise K-band local luminosity function.
In this paper, we will discuss
the imaging and photometry of the survey, and define a K-band galaxy
number-magnitude relation to K=16.
We also apply the no-evolution model
to the observed galaxy number-magnitude 
relation and discuss the distribution
the local K-selected galaxy density. 
In the following paper, we will analyze the morphological mix of the
K-selected bright galaxy sample up to K=14 (Huang \& Cowie 1996).\\

\begin{center}
{\bf 2. Observation and Data Analysis}\\
{\bf 2.1. Field Selection}\\
\end{center}
\vspace{5mm}

  The Hawaii wide-field K-band survey consists of two parts. The
first part was done in two relatively small areas, totaling about 1.58 square 
degrees, to a magnitude limit of K=14.5. This part of survey yielded a
relatively small sample whose redshifts could be easily measured.
The data analysis and redshift measurement are given in Gardner(1992)
and Songaila {\sl et al.} (1994). We will re-analyze these data
together with the second part of the survey.\\
   
  The second part of the survey, 
is designed to use the newly-developed multi-fibre spectrograph
(the 2DF) on the Anglo-Australian Telescope(Taylor 1994) for
followup redshift measurement.
Four fields, named SA, SB, SC and SD,  were 
selected on the equator, so as to be observable from both Mauna Kea, Hawaii
and the AAO, Australia.
These 4 fields avoid rich clusters and very
bright stars. We also selected fields on which
previous surveys have been carried out.
Glazebrook {\sl et al.} (1994) conducted a medium-deep
K-band survey on the fields SA, SC and SD with small sky coverage.
The fields SC and SD are also sites of the
Anglo-Australia Redshift Survey, 
each with a total of 14 square degrees of coverage
(Peterson {\sl et al.}  1986).
All four fields are at high galactic latitudes to avoid galactic
extinction. After we trim those parts of the areas with
insufficient S/N ratio, each field splits into 2-4 pieces
and the total number of areas becomes 13. 
The positions
and completeness limits in the K-band are summarized in Table 1.

\begin{center}
{\bf 2.2. Observation and Data Reduction}
\end{center}
\vspace{5mm}

  The observations were carried out on the University of Hawaii telescopes 
on Mauna Kea, over more than a hundred nights during the 1993-1995 period 
The K-band images were taken using the 24 inch telescope
with the NICMOS IR camera (256$\times$256 pixels). 
The field of view of the NICMOS camera on the
24 inch telescope is approximately 8 arcminutes with a scale of 
2.02 arcseconds per pixel.
Each field was observed frame by frame 
in the Dec. direction to form a one-degree strip, 
then each strip was stepped in the R.A. direction.
The offset between
frames in each strip is about 4 arcminutes, or half field of view.
Hence, a strip contains 16 images. The offset between each strip is also
4 arcminutes. Therefore, each pixel was exposed 4 times in total, except 
those on the edges and corners which were exposed 2 times 
and 1 time respectively.
These edges are trimmed in the final mosaic frames.
With a 2 minute exposure time for each single image,
the final mosaiced K-band image has an exposure time of 8
minutes. The advantage of this observational pattern is that
bad pixels and cosmic rays are easily removed, as we form the final image
by taking the median of the 4 counts on each pixel. Instead of using
a standard K-filter, we used the K' filter (Wainscoat and Cowie 1992), 
which is slight bluer than the standard K filter, to avoid
the thermal IR emission.
The K' filter significantly reduces the sky
background. The background of each frame with a 2 minute
exposure time is still very high, on average 3000-5000 counts.
The background also increases with airmass, so each  observation
was limited to an hour angle less than 2.5 hours.
There are further advantages
to the low airmass limitation: it ensures a uniform S/N ratio
throughout whole area; and it minimizes the zero point problem
due to different airmasses in the final mosaic, since the magnitude difference
for the background in our fields 
between zenith and hour angle of 2.5 hour is less than 0.01.
The near-infrared standard stars were taken from the list of
Elias {\sl et al.} (1982), with K magnitudes of $6<K<9$.
We used a linear relation derived by Wainscoat and Cowie (1992) to
transfer the K magnitudes of the stardard stars to the K' magnitudes. Since
the linear dynamic range of each pixel in the NICMOS camera is up
to 20000 count rates, the exposure time must be short enough for
the standard stars to prevent saturation. We took 5 seconds
exposure time for our standard stars. The highest peak count rates 
of our standard stars are about 12000.\\

  The image processing for the near-infrared images is
the same as that used in the Hawaii Deep Survey, described in Cowie
{\sl et al.} (1994).
We would like to emphasize some important points here again.
For the K-band images, the dark current is insignificant, and therefore
was not subtracted from each raw data frame. The flat field was
created by using the high sky background of the raw data. Due to the variation
of the sky OH emission, the sky flat field in K-band may 
change on time scales of
20$-$30 minutes (Glazebrook {\sl et al.}  1994). 
Hence, for each raw K band image, the flat field was generated by
using a median filter of ten nearby adjacent frames in the observation time
sequence. 
Since the sky background is so high,
the noise is then dominated by the background noise. 
The flat field generated from
ten raw frames should have a lower noise level than the expected final
mosaic images which are a median of only four raw images so that the flat field
created in this way does not contribute significant 
noise to the final image.\\

  We were successful in obtaining B and I band photometry of entire area
reported in Table 1 covered by the K-band exposures.
The observations in I and B band were primarily obtained  
on the 88 inch telescope
with a Tek2048 CCD. Of the total area for which we obatined the B-band
photometry, about 25\% was observed
with the 24 inch. The purpose of the optical imaging is to
separate stars from galaxies in the color-color diagram, and to study the
morphology of the K-selected galaxies. With a scale of 0.22 arcsecond per
pixel and average seeing of 0.8 arcsecond, the I-band images have
much higher spatial resolution than the K-band images. The field of view 
for the CCD images is about 7 arcminutes at the 88 inch. The observing pattern 
for the optical observations is similar to
that of the K-band observations. The effective exposure time is 2 minutes
for the I and B band images taken on the 88 inch telescope, and 20 minutes for
the B band images taken on the 24 inch telescope.\\

 The data reduction for the optical images follows standard procedures. 
Each raw data frame was bias subtracted, and then divided by a flat
field. The procedures for obtaining the flat fields for 
I and B band images are different.
For the I band
images, the flat field was created in the same way as for the K-band images.
Since the I-band sky-background is also high and stable, a sky flat 
was used in the I-band image processing.
Due to the extremely low sky background level in B band,
a twilight flat field was used for the B band images.\\

 The final registering of the K-band images 
is difficult. Due to the inaccurate pointing
of the UH 24 inch telescope, the offset between each neighboring frame needs
to be determined precisely. Lack of bright stars and the high noise level
in the primary frames can make this measurement difficult. We have developed
the following method to cope with this problem.
The offset between neighboring
frames was measured by using bright stars in the overlapping
area. Saturated stars were not used in measuring positions.
For those frames with no
usable bright stars but apparent faint objects above the noise, a
cross-correlation function was used in the overlapping area to determine
the offset. At this point, 
there were only 5\% of the frames left whose cross correlation
functions with their neighbor frames had too low S/N to determine
the offset.
The first mosaic was constructed without those frames.
The position of each frame on the final mosaiced frame 
was calculated from its offsets to its neighbor frames. The counts for 
each pixel of the final frame
was the median of counts of the 4 exposures, except for those
pixels on the edges. The S/N ratio for those fully covered pixels increases
by a factor of 2.
We roughly estimated the position for the remaining frames
by guessing their offsets to the neighbor
frames, usually  a half of the raw field of view
as designed by our observing stepping pattern; then cut out a similar
field of view from the first-pass mosaic, where the composite S/N
has generally been substantially improved over that of a single
neighboring ``raw'' data frames,
and then ran a cross-correlation function with
the corresponding frame to be incorporated into the final position. 
In this second-pass, the peak of the cross-correlation function
can be well determined, since one image has an improved S/N ratio usually
by a factor of $\sqrt3\sim1.71$. The position of this frame in the final
composite, therefore, could be located.  The frame was then incorporated
into the composite mosaic. Berfore each frame is added to the composite
mosaic, the peak of cross-correlation function between this frame and the
composite mosaic is measured again to supress the growth of errors by
random walks. This process was repeated until all the remaining
frames that had been omitted from our first-pass composite mosaic
were
located. Before the final mosaic was generated, 
an airmass correction for extinction was applied
to each of the original frames.\\

  The registering of the optical images is much easier, 
since the effective exposure
time for the I and B band observation results in images 
deep enough to have many usable
stars. Therefore the position of each CCD image can be determined precisely in
the final mosaiced image. The K and B band images are finally registered
to the I band images. Due to the large pixel of the 
K-band images($\sim$ 2.02'' pixel$^{-1}$),
the position error of a star between the K-band image and optical images 
may be as large as 2'', though the position error 
between the I and B band images is much smaller.
The final three color images were cut into pieces, each
with size $7.3'\times7.3'$ for easy storage and further analysis.\\

\begin{center}
{\bf 2.3. Detection and Photometry}
\end{center}
\vspace{5mm}
  
   The object selection was carried out on 
the registered K-band images. I and B
band counterparts can be easily located, allowing for positioning error, 
at the position where the K-band
objects are detected. Since the original low-resolution K-band images
were registered to I-band images which had much smaller pixel scales,
this procedure tends to smooth the K-band images, and no further
smoothing is applied. For the UH 24-inch telescope, there are
many factors contributing to the image quality, such as dome seeing,  
inaccuracy of tracking, and wind shake. The width of the 
Point-Spread-Function(PSF) on K-band images is slightly larger than
the pixel scale. Our detection procedure
is similar to Tyson's (Tyson 1988). The criterion for identifying an
object on the registered K-band images is  that an object
must cover an area of at least
12 arcsec$^2$ down to a K-band surface brightness of 20 mag arcsec$^{-2}$. 
This criterion is equivalent to having 3 connected pixels, each having
counts higher than 3$\sigma$ on the K-band images.\\
  
  The completeness of the sample was tested by using extensive
Monte Carlo simulations. We first mimiced the noise according to
the real images. Then five bright galaxies with different Hubble
types were selected. After rescaling the counts of these galaxies by
different factors, we add
the simulated noise and ran the detection process to
obtain the galaxy detectability as a function of magnitude.
We found that, in each field, such functions vary slightly according
to galaxy type. The sixth column of the Table 1 gives the median recovery
rate at K=16 mag for each field, ranging from
70\% to 97\%. \\

  The method of photometry is very important in analyzing our results, 
and therefore
should be treated very carefully. As many previous studies have indicated
that, for a faint galaxy, its isophotal photometry is 
strongly dependent on redshift and hard to
model, we have chosen to use aperture photometry.
Because of the low spatial resolution in the K-band images and the 
large angular size of the bright galaxies,
a large aperture is required.
An 8-arcsec-diameter aperture is adopted.
This aperture magnitude is corrected to 20 arcsec diameter.
For bright galaxies, specifically $K<13$, their
sizes are generally larger than 20 arcsec, and corrections of aperture
magnitudes for these galaxies are large and variable.
For these galaxies,  we used an isophotal
magnitude measuring the isophotal magnitude to a surface brightness
of 20 mag arcsec$^2$. 
Since these bright galaxies are local, there
is little cosmological effect on their isophotal magnitudes.\\

  Since stars and galaxies have different growth curves, their 
aperture magnitudes were corrected separately.
we first classified stars and galaxies by
using their morphological indicators measured in the I-band images,
which will be discussed in next chapter.
Since a bright galaxy is more extended than a faint galaxy,
we have to correct the magnitude differently according to its
apparent magnitude. 
Then we divided galaxies and stars into bins according to their magnitudes
using a 0.2 mag bin size. 
The growth curves were measured by using well-isolated stars and galaxies
in each of the bins. The star and galaxy corrections in each bin
were made from the
median of these stellar and galactic growth curves. 
Due to the variation of the PSF, the corrections for each field are treated
separatedly.
Fig. 1 is an example from one field, and it shows that the corrections for
galaxies are
a function of the apparent magnitude for galaxies. 
The corrections for stars are
not a function of aperture magnitude since they have 
same PSF at all magnitudes.\\

  For I- and B-band photometry of the K-selected sample
the magnitude was measured with the same procedures as
for the K magnitudes centered on the position
of the objects in that color.
Since we have a large aperture (8 arcsec), it
makes almost no difference whether we measure the optical 
magnitude at the position of the K-selected object or the best optical
position. 
If more than one object was found close to the K-selected 
object within 2 arcsec, 
the one closest to the position of the K-selected object was used.
If nothing was found, a 3$\sigma$ upper limit was adopted.
The optical photometry was performed in exactly the  same
way as the K-band photometry; except that 
on the I-band images, an extra parameter, 
the fourth moment of the radius, was measured on each object for morphological
classification. This parameter is defined as
\begin{equation}
      r_4=\frac{\int_0^Rf(r)r^5dr}{\int_0^Rf(r)rdr}
\end{equation}
where f(r) is the flux, and R is the maximum radius for the integral.  If R is
too large, $r_4$ can be dominated by noise. We found empirically that
the noise can be minimized for $r_4$ when R=3 arcsec.\\

  The same procedure of detection and photometry was run again through the
B-band images to obtain the B-band galaxy counts. The detection criterion
is still 3 connected pixels with counts $3\sigma$ higher than the noise.
This criterion transfers to the surface brightness as, 22.86 mag per square
arcsec for the images taken with the 24 inch and 24.21 mag per square
arcsec for the images taken with the 88 inch. Our Monte Carlo simulation
indicates that the completeness is at 20.5 mag for the images taken with
the 24 inch and at 22.2 for the images taken with the 88 inch. The
morphological index $r_4$ is calculated on the I-band countparter. 
For those galaxies with B magnitude deeper than 20, their K-band magnitudes
are generally beyond the limit of our detection. 
Hence, the star-galaxy classification
cannot be conducted on the color-color diagram, and the morphological
analysis is
the only way to discriminate between stars and galaxies. 
However, for those galaxies fainter than B=21 mag, their
$r_4$ becomes noisy, so we measure the B-band galaxy counts only 
to B=21 mag.\\ 

\begin{center}
{\bf 2.4. Star-galaxy classification}
\end{center}
\vspace{5mm}

   Star-galaxy classification was carried out in two ways:
using object morphology and color properties.
The $r_4$ as a morphological
indicator is very sensitive to the shape of an 
object. Stars all have the same $r_4$
since they all have the same PSF. 
The $r_4$ of an extended object, however, must be
larger than that of a PSF, since an extended object contributes more light
at large radius than does a PSF. However, it has one
disadvantage for morphological identification. As mentioned before,
$r_4$ has to be calculated over a high S/N area, so
compact galaxies can be misidentified as stars. A different
classifying method must then be applied. The color-color diagram has
been proved as a good method to discriminate between stars and 
galaxies (Gardner 1992), since on an I-K vs B-I color-color diagram,
stars are clearly separated from galaxies. As Gardner(1992) noted,
the separation line is $(B-I)-2.5(I-K)=-2.0$, We found that this line
separates our data well, see Fig.2. With these two methods, our final
criteria to
identify a galaxy is: that an object has $r_4 > (r_4)_{psf}$ or that an
object has $(B-I)-2.5(I-K)>-2$. Fig.2 shows that both methods are consistent
with each other, as almost all morphologically identified galaxies also
lie below the separation line in the color-color diagram.\\

\begin{center}
{\bf 3. K-band Galaxy Number Counts}\\
{\bf 3.1 Number Counts}\\
\end{center}
\vspace{5mm}

  The K-band galaxy number counts are shown in Fig.3. Due to
the different completeness limits in different fields, the counts on
each field were corrected separately, as described in
\S2.3. The final counts are the average of all fields weighted
by their areas. The number counts cover the magnitude range from K=12 to K=16.
The fitted logarithmic slope of our observed counts shown as filled circles
on the plot,
is $dlog(N)/dm=0.689\pm0.013$ (the Euclidean value is 0.6).
The data are also presented in the Table 2. The $1\sigma$ 
errors for the counts
in each bin are also shown, and a detailed discussion is given in \S3.2.
The average field-to-field variation among our 6 fields at each magnitude bin
is also listed in Table 2. We found that this variation has a minimum at
K=14.5 of only 5\%. At the bright end, the 
variation results from the Poisson noise.
However, at the faint end this variation may be caused by other factors as
discussed in the following text.

  In Fig.3, we also show the data from other surveys. Generally the agreement
between the surveys is excellent.
Our counts at K=11.75,
12.25 match the counts of Mobasher {\sl et al.}'s (1986) well. 
Between K=12.5 and
K=15 our counts are slightly higher 
than the other counts. In this magnitude range,
however, both the Hawaii medium-deep survey(HMWS)(Gardner 1992) and 
Glazebrook {\sl et al.}'s (1994)
survey have much larger error bars. We also note that between K=15 and
K=16 our counts are systematically higher than those of the HMWS. Though the 
difference is only about 10\%, within the 1$\sigma$ error bars 
of the HMWS,  this may also be due
to the difference between the two photometric systems used. We also note
that over the same magnitude range the counts of Jenkins \& Reid (1991), 
HMDS and HDS (Gardner 1992) are
all slightly higher than the present data, 
probably due to their large Poisson noise.
Recently, Gardner {\sl et. al.} (1996) have also completed  a wide-field
K-band survey. Their coverage is nearly 10 square degree, about the same
size as ours. Their counts in the range $13<K<16$ are within 1$\sigma$ of
our counts. However, their counts at the bright end of $K<12.5$ are higher
than ours. We argue that this difference is very likely due to the
small sample statistics.   
The comparison between 
the observed counts and model predictions will be given in \S3.4\\

\begin{center}
{\bf 3.2 Uncertainty in the Number counts} 
\end{center}
\vspace{5mm}

   Three factors contribute to the uncertainty in the galaxy number counts:
the Poisson noise, the galaxy distribution and the magnitude error. 
Understanding such uncertainties is the key
to modeling the galaxy number counts, especially in obtaining the normalization
factor $\phi_*$. We first analysize the uncertainty of the galaxy number counts
caused by the Poisson noise and the galaxy distribution.
Several authors (Glazebrook {\sl et al.}  1994, Djorgovski
{\sl et al.}  1995) have noted the contribution of galaxy clustering to
the uncertainty. However, the distribution of large scale structures
such as rich clusters and large voids are poorly known. We can only
model the uncertainty caused by the galaxy-galaxy correlation which
is well known.\\

  We first give the solution of the error caused by galaxy-galaxy 
correlation.
We approach this problem in non-relativistic form.  Since the galaxy
correlation scale is about $5h^{-1}$ Mpc, 
the non-relativistic approach gives a good approximation. Here, 5Mpc
is corresponding to an angular size of 0.8 degree at a redshift of
0.15, a characteristic redshift of this K-slected galaxy sample.
The uncertainty for a galaxy sample (Peebles 1975, 
Roche {\sl et al.}  1993, Ferguson \& Binggeli 1994) in one field 
can be written as
\begin{equation}
\sigma_i^2=N_i+2.24N_i^2A\Omega_i^{\frac{1-\gamma}{2}}
\end{equation}
where i means the {\sl i}th field, $\Omega_i$ is its area in unit of square
degree, A is the constant in an angular correlation function (Peebles 1980),
as
\begin{equation}
A=r_0^{\gamma}\sqrt{\pi}\frac{\Gamma(\frac{\gamma-1}{2})}{\Gamma(\frac{\gamma}{2})}\frac{\int_0^{\infty}x^{5-\gamma}p^2(x)dx}{[\int_0^{\infty}x^2p(x)dx]^2}
\end{equation}
$p(r)$ is a selection function, $r_0=5h^{-1}Mpc$, and $\gamma=1.77$.
For a sample with apparent
magnitude m,
\begin{equation}
p(r,m)=\phi(m-5log(r)-25)
\end{equation}
$\phi(M)$ is the luminosity function. We adopt the 
Mobasher {\sl et al.} (1993) K-band luminosity function here with
$M_*=-23.59+5log(h)$ and $\alpha=-1$.
For the observed galaxy number counts, each bin is a sample with
apparent magnitude m. Therefore, by applying equation (4) to (3) to calculate
the constant A, then putting (3) into (2), we obtain an
analytical solution of the uncertainty in observed 
galaxy counts at each magnitude;
\begin{equation}
\sigma_i^2=N_i(m)+5.3(\frac{r_0}{r_*})^{\gamma}\Omega_i^{\frac{1-\gamma}{2}}N_i^2(m)
\end{equation}
and $5log(r_*)=m-M_*-25$.
N(m) is the observed counts
in each bin. The final uncertainty for the number counts 
per magnitude per square degree in all fields is
\begin{equation}
\sigma(m)=\frac{\sqrt{\sum\sigma_i^2}}{\Omega \Delta m}
\end{equation}
with $\Delta m$ as the bin size and $\Omega$ as the total area 
($\Omega(K<14.5)=9.81 d^{-2}$ and $\Omega(K>14.5)=8.23 d^{-2}$).
Fig.4 shows the contribution of both Poisson noise and galaxy correlation
uncertainty to the total error in this survey. The
1$\sigma$ errors of the observed counts calculated by using above 
equations are listed in the third column of Table 2. It can be proved that
for a survey with small sky coverage, the uncertainty caused by galaxy 
correlation is much smaller than the Poisson noise. As we can see in
Table 2, the 1$\sigma$ is less than the field-to-field variation
at $K>15$, but comaprable to the field-to-field variation at $K<15$.
Hence, it is very likely that the other factors (rich clusters
and large voids) than the Poisson noise
mainly contribute to the field-to-field variation at the deep
magnitude ($K>15$) counts.\\

  Magnitude errors may also cause changes of the galaxy number
counts. These magnitude error result from two sources, systematic error
and random error. In our case, the systematic error is the error in
the zero point for the photometry, which in K-band was
$\sigma_s \sim 0.05 mag$ between observing runs. 
This error does not change the slope of log(N), but shifts the log(N)
in horizontal direction in the count-magnitude diagram. We can transfer
this magnitude error to the count error as
\begin{equation}
\frac{\sigma}{N(m)}=\alpha\sigma_s
\end{equation}
where $\alpha$ is the slope of Log(N(m)).
If we take the slope of 0.689 and $\sigma_s$ of 0.05, we obtain
$\sigma/N=3.5$\%, which is less than the field-to-field variation.
The final contribution of this systematic error to the galaxy counts
will be less than 3.5\%, since the final result is the average counts
of all observing runs.\\

  Random error in the magnitudes can also change log(N). In 
the K-band survey, the sky background counts are much higher than galaxies
counts. Hence the random error in the  magnitude of a galaxy is mainly caused
by the sky background noise. Our Monte Carlo simulation indicates that
the random error changes increasingly from 0.01 mag at K=13 mag to 0.2 mag
at K=16 mag. The observed profile of the counts as a function of magnitude
is a result of the real profile of the counts convolving with the random
error distribtuion function, which is usually a Gaussian distribution.
If the random error $\sigma_r$ is a constant, 
we can obtain an analytical solution
as
\begin{equation} 
\log(N)_{obs}=\log(N)_{real}+1.15\alpha^2\sigma_r^2
\end{equation}
However, we have know the $\sigma_r$ increases when
the K magnitude becomes faint. Though we can not work out an analytical
solution in this case, we can qualitatively analyze this problem.
As both random magnitude error and intrinsic galaxy counts increase
with increasing magnitude, on average  there are more fiant galaxies
erroneously appearing in a brighter bin than brighter galaxies erroneously
appearing in a fainter bin. Hence, the observed slope of $\log(N)$ becomes
steeper than the real one. This trend can also be seen in above equation.
As we consider the slope change in the magnitude range from K=13 to K=16,
this range is much larger than the random magnitude error $\sigma_r=0.2$. By
considering that only the galaxies at nearby bins can contribute
to erroneously increasing of the galaxy counts, we argue that it is a 
good approximation to use above equation to estimate the change of the slope
as
\begin{equation}
\Delta\alpha=1.15\alpha^2\frac{\sigma_{r1}^2-\sigma_{r2}^2}{m_1-m_2}
\end{equation}
As a first order approximation, we use $\alpha=0.689$, $\sigma_{r1}=0.2$ at
$m_1=16$ and $\sigma_{r12}=0.01$ at $m_2=13$ to obtain that 
$\Delta\alpha=0.008$ and $\Delta N/N=5.1$\% at K=16. 
Since the 1$\sigma$ of the slope
from fitting is 0.013, this effect does not change 
the slope of the counts significantly.\\

\begin{center}
{\bf 4 B-band Galaxy Number Counts}\\
\end{center}
\vspace{5mm}
  
  The B-band galaxy counts are presented in Fig.5, togather with the
data from other surveys. The data are also listed in Table 3.
The purpose of measuring the B-band counts is
to monitor our fields. Though we carefully selected the fields, we
still need to know if there are any special galaxy distribution
structures in our fields to distort the galaxy counts. 
The B-band galaxy counts have been studied substantially
, as sammaried by Koo \& Kron (1992). In the B-band magnitude range
(14 - 21 mag) which our counts cover, some survey areas cover as much as
4300 square degrees (Maddox {\sl et. al.} 1990).  Our B-band galaxy
counts are consistent with the counts of other B-band surveys including
Maddox {\sl et. al.}'s survey. This consistency means that our fields
are an average field, and the K-band galaxy counts measured from
our fields may not be distorted. The physical relation between
the K-band counts and the B band counts will be discussed later.\\

  The $1\sigma$ error of the B-band counts listed in Table 3 are
calculated by using equation 5 and 6 with a set of parameters of a 
B-band luminosity function. Like the K-band counts, this error is
also dominated by the Poisson error. The field-to-field variations
for the B-band counts are listed in the third column of Table 3. 
Our zero point error for the B-band photometry is about 0.03 mag,
and the slope of the B-band counts in the range of 15 mag to 21 mag
is 0.38. By using equation 7, this systematic error transfer an error 
of the counts at about 1\% level. The random error of the B
magnitude is different from that of the K magnitude. Since the B-band
sky background is very low, the random error is mainly caused
by the photo counting statistics of a galaxy, and is inversely 
proportional to square root of its counts. Hence, the faint galaxies
have larger magnitude errors than the bright galaxies. 
The random error is only about 0.01 at B=21 mag, and negligible at B=16
mag.  Putting this error and the slope of the B-band counts in
equation 9, we find that the steepening effect of the random magnitude
is negligible for the B-band counts.\\

\begin{center}
{\bf 5 Modeling the K-band Counts}\\
{\bf 5.1 No-Evolution Model}\\
\end{center}
\vspace{5mm}

  The K-correction is also a key issue in understanding a faint galaxy's
magnitude. In the optical bands, the K-correction is a strong function of
morphological type (Cowie {\sl et al.}  1994). 
Thus, in fitting no-evolution models to optical counts, usually 
measured in the B-band
(King \& Ellis 1985, Yoshii \& Takahara 1988),
the mix of morphological type has to be determined. We also have to
obtain the type-dependent luminosity functions, which are usually assumed
to be the same as for the observed total luminosity function. The
no-evolution modeling for the 
optical counts is, therefore, highly dependent on
the mix of morphological types. However, the K-correction in the K-band
is different. 
Firstly the near-infrared spectrum of a galaxy
is flater and more featureless than in optical band
so the K-band corrected magnitudes are consequently
more reliable than the optical corrected magnitudes. Secondly,
the K-correction in the K-band is only a weak function of 
morphological type. Up to
z=1, to a good approximation, we can ignore the differences of 
K-correction in the K-band among different 
Hubble types (Glazebrook {\sl et al.}  1994).
The details of the K-band K-correction, however, are poorly determined.
Currently the K-band K-correction is derived either from evolving
galaxy models (Rocca-Volmerage and Guideroni 1988, Bruzual and Charlot,
1993, and Buzzoni 1995) summarized by Glazebrook {\sl et al.} (1995a),
or from interpolation of the J-K and H-K colors (Cowie {\sl et al.} 
1994). The K-corrections derived from these methods, as we see in Fig.6,
do not agree with each other exactly. Considering the uncertainties
of these methods, however, the differences are reasonable.
As seen in Songaila {\sl et al.} (1994), even 
at their deepest magnitude (K=20) 
the median redshift
of the K-selected galaxies is still substantially 
less than one. Since our number-counts are below K=16, 
we can construct a no-evolution model for our K-band
counts with a general K-band luminosity function and a simple
K-correction term. \\

  The traditional K-band luminosity
function (Yoshii \& Takahara 1988) is derived in this way:  
first by assuming that each Hubble
type has a uniform color, B-K; then by converting each optical 
type-dependent luminosity function into K-band type-dependent 
luminosity function
according to color; and finally by adding all type-dependent luminosity
functions weighted according to the type mix. More directly,
we can just use the observed K-band luminosity function.
Unfortunately, the K-band luminosity functions reported in the literature
do not agree with each other well. Mobasher {\sl et al.}(1993)
conducted a redshift survey on a K-selected subsample from the B-selected
sample of the Anglo-Australia Redshift Survey, and constructed the
first K-band luminosity function fit to a Schechter function
with $M_*$=-23.59($h=1$ hereafter) and
$\alpha=-1$. The K-band
luminosity function which Glazebrook {\sl et al.}(1995a) obtained
directly from their K-band survey
has a similar shape, but their $M_*$ is significantly fainter,
$M_*=-22.75$. The most recent determination of 
Cowie {\sl et al.} (1996) finds an $M_*$ consistent
with that of Mobasher {\sl et al.} (1993) and a similar shape. Both
luminosity functions will be
used in constructing the no-evolution model.\\

  The no-evolution model can also be presented in another way, as a
mean color-magnitude relation. Specifically, we will derive $<I-K>$ 
as a function of K magnitude. Due to the difference of 
the I-band K-correction for different spectral type galaxies,
the $<I-K>$ of each type of galaxy has to be treated
separatedly. Since we do not have the K-band type-dependent luminosity function
we have to adopt the general K-band
luminosity function for  each type galaxies. To be consistent
with the I-band K-correction in Cowie {\sl et. al}'s
paper (1994), we simply divide galaxies into E/SO, Spiral, Irregular
galaxies, and adopt the I-band K-correction directly from their paper.
Then, for each classified type galaxies, we have
\begin{equation}
<I-K>=<I-K>_{z=0}+\frac{\int n(z,m) (kc_i(z)-kc_k(z))dz}{\int n(z,m)dz}
\end{equation}
where n(z,m) is the no-evolution model of the galaxy counts
as a function of redshift and magnitude,
$kc_i$ and $kc_k$ are the I and K band k-correction terms. In Fig.7,
the models of $<I-K>$ for E/SO, Spiral, Irr are plotted in 
comparison with the observed data.
Each model has been normalized to the colors of the bright galaxies
whose morphologies in their CCD images are classified by authers.
The models with faint $M_*$ predict bluer colors at the faint magnitude
end than those with bright $M_*$.  This is because, for a flux limited
sample, a model with faint $M_*$ predicts fewer distant galaxies than
a model with bright $M_*$, and the I-band K-correction makes the color 
of distant galaxies redder than those of nearby galaxies.
The observed $<I-K>$ lies between the models of E/SO and Spiral galaxies
with the same trend as those of the models.
This indicates that the galaxies in the K-slected sample are mainly E/SO
and early type Spiral galaxies, and a no-evolution model is good enough to fit
the observed color.
To combine our models to fit the observed
data suggests that, in the K-selected galaxy sample, about 50\% of them are
E/SO, another 50\% are Spiral galaxies, and there are very 
few irregular galaxies.\\

\begin{center}
{\bf 5.2 Local K-selected Galaxies}
\end{center}
\vspace{5mm}

  Understanding the local galaxy sample is also essential 
in connecting the model to the observed
counts, since the local galaxy sample serves as a 
reference point for the models. 
We selected a subsample with a limiting apparent magnitude of K=15 mag.
If we also add the counts of Mobasher {\sl et al.} (1986), the local
counts cover from K=10 to K=15.
For such a magnitude range, a no-evolution model can be characterized 
by its slope. We have calculated the slopes of no-evolution models
with different luminosity functions, different geometries, and different
K-corrections. The slope of a
no-evolution model in $10<K<15$ is independent on $q_0$, and 
varies only from 0.605 to 0.623, for all current varieties of 
$M_*$ and the K-correction. The result is summarized in Table 4 .
We conclude that the slope is very weakly dependent on the $M_*$, 
and the difference in the current K-corrections does not 
significantly change the no-evolution model. 
The slope of the observed counts at this
magnitude range, however, is much larger than the maximum slope 
of the theoretical
prediction. As shown in Fig.8, the slope of the counts in this survey
is $0.69\pm0.014$. 
When we combine our counts with 
those of Mobasher {\sl et al.}'s (1986), the slope becomes 
$0.68\pm0.012$. Gardner {\sl et al.} (1996) obtained a flater slope,
as 0.63 in $10<K<16$. However, by excluding their bright counts measured
with only a few galaxies, their
slope in $13<K<15$ is 0.65,  which is not as steep as ours. 
It is possible that their
fields contain more bright galaxies than the average, 
since their B-band counts below $B=18$ are
significantly higher than the B-band counts of Maddox {\sl et al.} (1990).
By combining our counts with Gardner {\sl et al.}'s (1996), we still
obtain a much steeper slope of 0.67 in $13<K<15$.
Hence the no-evolution model slopes are $5\sigma
-7\sigma$ away from the observed slope.
Normalizing our no-evolution model with the steepest slope to
the counts at K=10.25 as in Fig.8, we find that
the model can not fit the observed counts. The result is independent of
our selection of luminosity functions, normalizations and geometries. 
The only parameter that determines the slopes of no-evolution models
is the K-correction. Though the K-corrections derived from various methods are
quite consistent with each other as we indicated above, they could be still
poorly determined. Near-infrared emission features could be missed
in the stellar synthesis models and the color interpolation.
The features in between $2.0\mu m$ to $2.2\mu m$, such as He~I, H$_2$ S(1),
and Br~$\gamma$ could change the K-correction at low redshifts, and therefore
change the slope of a no-evolution model in $10<K<15$. We found that we could
fit the number counts
by artificially increasing the K-correction by a factor of 3 at between
z=0 and 0.2. 
However, we argue that such a large correction(more than a magnitude) 
is very unlikely
to be the case. Ridgway {\sl et al.} (1994) took near-infrared spectroscopy
on their infrared-luminous galaxy sample. Most of the galaxy spectra 
in the range of $2.0\mu m$ to $2.2\mu m$ are featureless, even if
where the infrared-luminous galaxies are gas-rich galaxies. Our K-selected
sample is dominated by early type galaxies, and cannot have such strong 
emission features. Hence we do not think that the current K-corrections
 can deviate from the real one by such a large factor.
The difference between
the observed counts and the model prediction at $13<K<15$ implies
either galaxy evolution at low redshifts or a local deficiency of galaxies.\\

  Galaxies with apparent magnitudes of $13<K<15$ are located at $z<0.2$
(Songaila {\sl et al.}  1994). 
Galaxy evolution at such low redshifts 
was first proposed based on the B-band counts
(Maddox {\sl et al.}  1990). Maddox {\sl et al.} found that there was an
excess in their B-band
counts at $18<B<20$ over a no-evolution model. Their
magnitude range in the B-band is equivalent to $13<K<15$ in
the K-band.
Such galaxy evolution, however, is very unlikely
to be luminosity evolution since many redshift surveys (Cowie
{\sl et al.}  1996, Ellis {\sl et al.}  1996, 
Glazebrook {\sl et al.}  1995a, Songaila {\sl et al.}  1994) in both
B-band and K-band have shown that galaxies with luminosities brighter
than $L_*$ ceased evolving long before z=0.5.  The galaxy evolution at low
redshifts
suggested by some of these authors occurs only among 
low luminosity galaxies. By analyzing K-band and B-band luminosity
functions at different z (Cowie {\sl et al.}  1996, Ellis {\sl et al.}  1996),
they found that there were more dwarf galaxies in the past than now.
Cowie {\sl et al.} (1996) further suggest that higher density of dwarf galaxies
in the past is due to the evolution of the formation process.
However, this evolution is not significant at $z<0.2$.
We will be able to confirm this argument only after we compare the
luminosity functions at z=0 and at z=0.2, constructed from this sample
in the followup redshift survey.\\

  A local deficiency of galaxies seems the best explanation 
of the steep counts at the bright-end. 
This possible explanation has been proposed based upon the B-band counts
(Shanks 1989 and Driver {\sl et al.} 1995) and radio source counts
(Windhorst {\sl et al.}  1990, Condon 1989).
The model requires
that the Galaxy happens to be in an extremely large low-density region
in the universe with a scale from $300h^{-1}$ Mpc 
to $600h^{-1}$ Mpc (corresponding
to $z=0.1\sim0.2$). The galaxy number density of our neighborhood is then lower
than the galaxy number 
density at $z=0.1\sim0.2$ which may be  the true average galaxy number 
density of the universe. The difference between the observed counts
and the no-evolution model at $13<K<15$ 
normalized at K=10.25 then represents the
difference between the density of our neighborhood and the average
density of the universe. We find that the following function of $\phi_*(z)$
can increase the slope of a no-evolution model at $K<15$ to 0.69 (see Fig.8);
\begin{equation}
\phi_*(z)=\phi_*(1+exp(-(z/0.1)^2))^{-1}
\end{equation}
This function is almost constant at $z>0.2$, and so does not significantly
change a no-evolution model at $K>16$.
This density profile is very arbitrary, 
since the observed galaxy counts provide little constraint.
However, this function should also fit the B-band bright counts if
there is a real local deficiency. From the B-band bright galaxy
surveys (Shanks 1989 and Maddox {\sl et al.}  1990), we already know
that the slope of the B-band bright counts below B=20 
is steeper than that of a no-evolution model. 
Fig. 5  shows the observed B-band counts (including ours) and the no-evolution
model made by Ellis (1987).
Furthermore, our local hole model predicts that
the galaxy density at z=0.2 is twice  the local galaxy density at z=0.
When Shanks (1989) normalized his B-band no-evolution model at B=18,
the model indeed over-predicted the local galaxy density by a factor of 2.
Maddox {\sl et al.} (1990) also found that their B-band counts
at B=20 had an excess over the no-evolution model by a factor of
2. As many redshift surveys indicate, galaxies with $B=18\sim20$
are located at about z=0.2. Thus, the B-band counts show a
similiar density profile to that of the K-band counts. This
indicates a consistency for the local deficiency model. 
However, since
we also do not know exactly where galaxies stop evolving, a redshift
survey on a very large bright sample, such as this large
K-selected sample, is required to distinguish
these two effects on the steep counts at bright-end. Such a redshift
survey will be large enough to provide precise luminosity
functions at different redshifts. If the shape of K-band luminosity functions
in this redshift range remains the same, but $\phi_*$ changes with z,
this will confirm that
the large scale structure produces the steep slope at the bright
end of the observed counts.\\

  Voids and superclusters are the largest structures ever found
in the universe (Rood 1988, Bahcall 1988). Their scales are usually
from $20h^{-1}$ Mpc to $120h^{-1}$ Mpc. Since the previous
large redshift surveys have limited depth, no structure with a scale
larger than $200h^{-1}$ Mpc has been yet discovered. The presence of 
a local low-density region with a scale larger than $300h^{-1}$
Mpc must have a fundamental effect in our measurement of
$H_0$ and $\Omega$ (Turner {\sl et al.} 1992). It
implies that local measurements on smaller scales
may overestimate the ``true'' value of $H_0$ by as much as 33\%
and underestimate $\Omega_0$ by a factor of 2. If true, this would go far to
resolve current time scale problem at the expense of introducing
extraordinarily large scale inhomogeneity into the cosmological model.\\
 
\begin{center}
{\bf 6.Conclusion}
\end{center} 
\vspace{5mm}

  We have completed a K-band field galaxy survey with a total coverage of
about 10 square degrees. The results are summarized below.\\

  (1) The K-band counts from this survey extend from K=12 to K=16.
They are well matched to the counts of other surveys in this range, 
but have much smaller error bars than previous studies.\\

  (2) For a pencil-beam survey, 
the error caused by galaxy-galaxy correlation is smaller
than the Poisson noise. At the faint end of our counts, the error
is dominated by the large scale structures.\\

  (3) We construct no-evolution models for the 
K-band counts by using a variety of
luminosity functions, geometries and K-corrections. In the range of
$10<K<15$, these models yield very similar predictions.\\

  (4) The shape of the bright-end counts does not fit no-evolution
models. We find that none of the
theoretical and observational uncertainties that we have been
able to identify can cover this discrepancy.
We, then, conclude that there are changes in the galaxy luminosity
function at low redshifts for a K-selected sample.
The steep slope may be due to  a change in
$\phi_*$ with redshift, 
which reflects a substantial local deficiency of galaxies over scale
sizes of 300$h^{-1}$ Mpc. This would imply that the unvierse is
inhomogeneous on extremely large scales and that local measurements
overestimate $H_0$ by factors of up to 33\% and underestimate $\Omega_0$
by a factor of roughly 2. The conclusion 
needs to be confirmed by the follow-up redshift survey.\\

  We would like to thank Richard Ellis and Karl Glazebrook for much
help and advice with this project, and for their detailed comments
on the early drafts of this paper. We also thank the referee, Richard
Kron, whose comments greatly improved the paper.\\
\newpage
\begin{center}
{\bf References}\\
\end{center}
\vspace{5mm}
\hspace{-10mm}Bahcall, N. A. 1988, ARA\&A, 26, 631\\
\vspace{2mm}
\hspace{-10mm}Buzzoni, A. 1995 ApJS, 98, 69\\
\vspace{2mm}
\hspace{-10mm}Bruzual, G. A. \& Charlot, S. 1993, ApJ, 405, 538\\ 
\vspace{2mm} 
\hspace{-10mm}Condon, J. J. 1989, ApJ, 338, 13\\
\vspace{2mm}
\hspace{-10mm}Cowie, L. L., Gardner, J. P., Hu, E. M., Songaila, A., Hodapp, K.-
W., \& Wainscoat, R. J. 1994, ApJ, 434, 114\\
\vspace{2mm}
\hspace{-10mm}Cowie, L. L., Songaila, A., \& Hu, E. M. 1996, AJ, Submitted\\
\vspace{2mm}
\hspace{-10mm}Djorgovski, S., Soifer, B. T., Pahre, M. A., Larkin, J. E., Smith,
 J. D., Neugebauer, G., Smail, I., Mathews, K., Hogg, D. W., Blandford, R. D., C
ohen, J., Harrison, W., \& Nelson, J. 1995, ApJ, 438, L13\\
\vspace{2mm}
\hspace{-10mm}Driver, S. P., Windhorst, R. A., \& Griffiths, R. E. 1995, ApJ,
453,48\\
\vspace{2mm}
\hspace{-10mm}Elias, J. H., Frogel, J. A., Mathews, K., and Neugebauer, G. 
1982, AJ, 87, 1029\\
\vspace{2mm}
\hspace{-10mm}Ellis, S. R. 1987, Observational Cosmology, Proc. IAU Symp. No.
124, eds Hewitt, A., Burbidge, G. R. \& Fang, L. Z., Reidel, Dordrecht\\
\vspace{2mm}
\hspace{-10mm}Ellis, S. R., Colless, M., Broadhurst, T., Heyl, J., \& Glazebrook
, K. 1996, MNRAS, Submitted\\
\vspace{2mm}
\hspace{-10mm}Ferguson, H. C. \& Binggeli, B. 1994, Astro. Astrophy. Rev., 6, 67
\\
\vspace{2mm}
\hspace{-10mm}Gardner, J. P. 1992, Ph.D. Thesis, University of Hawaii\\
\vspace{2mm}
\hspace{-10mm}Gardner, J. P., Cowie, L. L., \& Wainscoat, R. J. 1993, ApJ 415 L9
\\
\vspace{2mm}
\hspace{-10mm}Gardner, J. P., Sharples, R. M., Carrasco, B. E., \& Frenk, C. S.,
 1996, MNRAS, in press.\\
\vspace{2mm}
\hspace{-10mm}Glazebrook, K., Peacock, J. A., Collins C. A., \& Miller, L., 1994
, MNRAS, 266, 65\\
\vspace{2mm}
\hspace{-10mm}Glazebrook, K., Peacock, J. A., Miller, L., \& Collins C. A., 1995
a, MNRAS, 275 169\\
\vspace{2mm}
\hspace{-10mm}Glazebrook, K., Ellis, R. S., Colless, N. M., Broadhurst, T. J., A
llinton-Smith, J. R., \& Tanvir, N. R. 1995b, MNRAS, 273, 157\\
\vspace{2mm}
\hspace{-10mm}Heydon-Dumbleton, N., Collins, C. A., \& McGillivary, H., 1989,
MNRAS, 238, 379\\
\vspace{2mm}
\hspace{-10mm}Huang, J.-S., \& Cowie, L. L. 1996 in press\\
\vspace{2mm}
\hspace{-10mm}Jenkins, C. R. \& Reid, I. N. 1991, AJ, 101, 1595\\
\vspace{2mm}
\hspace{-10mm}Koo, D. \& Kron, R. 1992, ARA\&A, 30, 613\\
\vspace{2mm}
\hspace{-10mm}Loveday, J., Efstathiou, G., Peterson, B. A., \&  Maddox, S. J. 19
92, ApJ, 390, 338\\
\vspace{2mm}
\hspace{-10mm}Lilly, S. J., Cowie, L. L., \& Gardner, J. P. 1991, ApJ 369, 79\\
\vspace{2mm}
\hspace{-10mm}Maddox, S. J., Sutherland, W. J., Efstathiou, G., Loveday, J., \&
Peterson, B. A. 1990, MNRAS, 247, 1p\\
\vspace{2mm}
\hspace{-10mm}Mcleod, B. C., Bernstein, G. M., Rieke, M. J., Tolletrup, E. V., a
nd Fazio, G. G. 1995, ApJS, 96, 117\\
\vspace{2mm}
\hspace{-10mm}Mobasher, B., Ellis, R. S., \& Sharples, R. M. 1986, MNRAS, 223, 1
1\\
\vspace{2mm}
\hspace{-10mm}Mobasher, B., Sharples, R. M., \& Ellis, R. S. 1993, MNRAS, 263, 5
60\\
\vspace{2mm}
\hspace{-10mm}Peebles, P. J. E. 1975, ApJ, 196, 647\\
\vspace{2mm}
\hspace{-10mm}Peebles, P. J. E. 1980, `The Large Scale Structure of Universe', P
rinceton University Press, Princeton\\
\vspace{2mm}
\hspace{-10mm}Peterson, B. A., Ellis, R. S., Efstathiou, G., Shank, T., Bean, A.
 J., Fong, R., \& Zou, Z.-L. 1986, MNRAS, 221, 233\\
\vspace{2mm}
\hspace{-10mm}Ridgway, S. E., Wynn-Williams, C. G., and Becklin, E. E. 1994,
ApJ, 428, 609\\
\vspace{2mm}
\hspace{-10mm}Rocca-Volmerange, B. and Guiderdoni, B. 1988, ApAS, 75, 93\\
\vspace{2mm}
\hspace{-10mm}Roche, N., Shanks, T., Metcalfe, N., \& Fong, R. 1993, MNRAS, 263,
 360\\
\vspace{2mm}
\hspace{-10mm}Rood, H. J. 1988, ARA\&A, 26, 245\\
\vspace{2mm}
\hspace{-10mm}Shanks, T., Stevenson, P. R. F., Fong, R., \& McGillivary, H. T.,
MNRAS, 206, 767\\
\vspace{2mm}
\hspace{-10mm}Shanks, T. 1989, In Galactic and Extragalactic Background
Radiation, Proc. IAU Symp. No. 139, eds Bowyer, S. \& Leinert, C., 
(Dordrecht:Kluwer), 269\\
\vspace{2mm}
\hspace{-10mm}Songaila, A., Cowie, L. L., Hu, E. M., \& Gardner, J. P. 1994, ApJS, 94, 461\\
\vspace{2mm}
\hspace{-10mm}Taylor, K. 1994, in Wide Field Spectroscopy and the Distant Univer
se, Ed: S. J. Maddox \& A. Aragon-Salamanca (Singapore: World Scientific), P15\\
\vspace{2mm}
\hspace{-10mm}Turner, E. L., Cen, R., \& Ostriker, J. P. 1992, AJ, 103, 1427\\
\vspace{2mm}
\hspace{-10mm}Tyson, J. A. 1988, AJ, 96, 1\\
\vspace{2mm}
\hspace{-10mm}Wainscoat, R. J. \& Cowie, L. L. 1992, AJ, 103, 332\\
\vspace{2mm}
\hspace{-10mm}Windhorst, R. A.,  Mathis, D. F., \& Neuschaefer, L. W.
1990, in ASP Conf. Ser. Vol. 10, Evolution of the Universe of Galaxies,
ed. Kron R. D.(Provo, UT:BookCrafters, Inc), 389\\
\vspace{2mm}
\hspace{-10mm}Yoshii, Y. \& Takahara, F. 1988, ApJ, 326, 1\\
\newpage
\begin{center}
{\bf FIGURE LEGENDS}\\
\end{center}
\vspace{20mm}
{\bf Figure 1} This is an example of the magnitude correction on one of
our fields.
We correct the aperture magnitudes of faint objects ($K>13$) 
to 20 arcsec diameter. The correction $\Delta M$is a function of the
aperture magnitude for galaxies, byt not for stars.  The solid line shows
the correction for galaxies, and the dashed line for stars.\\
{\bf Figure 2} This is an example of star-galaxy classification in  one field.
The star-galaxy Classification is performed in the (B-I) vs (I-K) color-color
diagram. The line marks the bounding color criterion for separating stars 
and galaxies 
(Gardner, 1992). The cross mark represents an objects with $r_4>(r_4)_{psf}$,
and the dot mark represents an objects with $r_4=(r_4)_{psf}$. The diagram
shows that morphological and color criteria match very well.\\
{\bf Figure 3} The filled dots are the counts from this survey.
GAR: Gardner {\sl et al.}, (1996);
MCL: McLeod {\sl et al.}, (1995);
HDS: Hawaii Deep Survey (Gardner, 1992); HMDS:  Hawaii Medium 
Deep Survey (Gardner, 1992); HMWS:  Hawaii Medium Wide  Survey (Gardner, 1992);
JR: Jenkins and Reid (1991); GLZ: Glazebrook {\sl et al.} (1994); DJO:
Djorgovski {\sl et al.} (1995); MOB: Mobasher {\sl et al.} (1986). 
The d(log(n))/dm=0.69 for the counts from this survey. The error bars of our
counts are calculatde by using equation 5 and 6 with a K-banb luminosity 
function\\
{\bf Figure 4} The errors in our number counts 
are calculated from Equation (6). It is shown that
the error caused by the galaxy-galaxy correlation
is smaller than the Poisson error. We have
$\sigma^2=\sigma_{poisson}^2+\sigma_{cluster}^2$, then, for a pencil-beam
survey, the clustering error is not important.\\
{\bf Figure 5}  Our B-band galaxy number counts are ploted with the
no-evolution model (Ellis, 1987). The counts of 
other surveys plotted here are those of
Gardner {\sl et. al.} (1996) (10$deg^{-2}$), Shanks {\sl et. al.} (1984), 
Heydon-Dumpleton {\sl et. al.} (1989) (100$deg^{-2}$), 
and Maddox {\sl et. al.} (1990) (4300$deg^{-2}$). The error bars of our
counts are calculated by using Equation 5 and 6 with a B-band luminosity
function\\
{\bf Figure 6}  The K-corrections are plotted. The Solid line is the
mean K-correction of all type galaxies from the JHK interpolation;
the dotted line is the k-correction
from Bruzual and Charlot (1993); the dashed line is the K-correction from
Rocca-Volmerange and Guiderdoni (1988). The K-band K-corrections
of the different type galaxies are very similar at low redshifts. Therefore
we only plot their means here.\\
{\bf Figure 7}  The left panel is the plot of I-K vs K. In the right panel,
the models of $<I-K>$ are plotted with the observed data. The upper lines
are the models of E/SO galaxies; the middle lines are the models of Spiral
galaxies; and the lower lines are the models of Irregular galaxies.
The solid lines are the models with the luminosity function of Mobasher
{\sl et. al.} (1993), and the dashed lines are the models  with the luminosity
function of Glazebrook {\sl et. al.} (1995a). The difference of these two
models is discussed in the text.\\
{\bf Figure 8}  The K-band galaxy counts are fitted by a pure
no-evolution model (solid line) and a no-evolution 
model with a density gradient (dashed line).
It is clear that the slope of a pure no-evolution model is very
different from that of the observed counts. The crosses are the counts
of Mobasher {\sl et. al.} (1986); the filled dots are the counts of this 
survey; and the triangles are the counts of Gardner {\sl et. al.} (1996).
Our fitting does not include the counts of Gardner {\sl et. al.} (1996).
However, their counts match our model very well in $13<K<15$.
\newpage
\renewcommand{\baselinestretch}{1}
\begin{center}
{\bf Table 1: Position and Completeness of Fields}\\
\begin{tabular}{l|l|l|l|c|l}
\hline
\hline
name & $\alpha$(1950) & ~$\delta$(1950) & ~{\sl b} & Area($d^2$) & $P_{16}$ $^{a
}$\\
\hline
SSA13$^b$ & 13:10:01.7 & ~43:00:33 & ~74 & 0.84 & $-$ \\
SSA17$^b$ & 17:04:59.8 & ~43:59:35 & ~37 & 0.74 & $-$ \\
SA        & 22:42:57.6 & -00:28:12 & -49 & 0.46 & 0.97 \\
SA        & 22:41:00.9 & -00:27:40 & -49 & 0.15 & 0.97 \\
SA        & 22:39:10.8 & -00:29:23 & -49 & 0.84 & 0.72 \\
SB        & 03:43:32.3 & -00:30:21 & -41 & 0.34 & 0.92 \\
SB        & 03:38:57.7 & -00:29:40 & -41 & 1.63 & 0.92 \\
SB        & 03:43:55.0 & ~00:26:49 & -41 & 0.14 & 0.70 \\
SB        & 03:42:39.8 & ~00:29:19 & -41 & 0.35 & 0.72 \\
SC        & 10:41:23.8 & -00:27:46 & ~49 & 1.72 & 0.96 \\
SC        & 10:43:29.7 & ~00:28:49 & ~49 & 0.30 & 0.96 \\
SD        & 13:41:56.3 & -00:25:49 & ~60 & 0.82 & 0.96 \\
SD        & 13:37:58.9 & -00:25:46 & ~60 & 0.74 & 0.96 \\
SD        & 13:43:11.8 & ~00:28:52 & ~60 & 0.32 & 0.96 \\
SD        & 13:42:10.2 & ~00:28:56 & ~60 & 0.42 & 0.96 \\
\hline
\end{tabular}
\end{center}
\begin{flushleft}
\vspace{2mm}
\hspace{-5mm}{\em a}: $P_{16}$ is the recovery rate of a galaxy with K=16.\\
\vspace{2mm}
\hspace{-5mm}{\em b}: These two fields are from Gardner(1992) and Songaila {\sl et al.}(1994)
with completeness at K=14.5.\\
\end{flushleft}
\newpage
\begin{center}
{\bf Table 2: K-band Galaxy Counts}\\
\begin{tabular}{c|c|c|c}
\hline
\hline
K & N $^a$ & $\sigma$ $^b$ & $\sigma_{ff}(\%)^c$\\
\hline
11.25 & ~~~~0.2 & ~~0.2& ~ \\
11.75 & ~~~~1.0 & ~~0.5& ~ \\
12.25 & ~~~~2.5 & ~~0.7& 68\\
12.75 & ~~~~7.1 & ~~1.3& 36\\
13.25 & ~~14.5 & ~~1.9& 41\\
13.75 & ~~30.0 & ~~2.7& 18\\
14.25 & ~~65.4 & ~~4.2& 10\\
14.50 & ~~99.2 & ~~7.6& 5\\
14.75 & 157.5 & ~~9.8& 7\\
15.00 & 224.1 & 11.8& 13\\
15.25 & 321.4 & 14.3& 11\\
15.50 & 495.9 & 18.5& 21\\
15.75 & 634.9 & 21.1& 10\\
16.00 & 825.1 & 24.3& 15\\
\hline
\end{tabular}
\end{center}
\begin{flushleft}
\vspace{2mm}
\hspace{-5mm}{\em a}: N is in unit of $mag^{-1}$ $d^{-2}$.\\
\vspace{2mm}
\hspace{-5mm}{\em b}: $\sigma$ is calculated by using Eq. (5) and (6).\\
\vspace{2mm}
\hspace{-5mm}{\em c}: $\sigma_{ff}$ is the ratio of the field-to-field
variation to the number counts.\\
\end{flushleft}
\newpage
\begin{center}
{\bf Table 3: B-band Galaxy Counts}\\
\begin{tabular}{c|c|c|c}
\hline
\hline
B & N $^a$ & $\sigma$ $^b$ & $\sigma_{ff}(\%)^c$\\
\hline
15.0 & ~~~~0.5 & ~~0.2& 20 \\
15.5 & ~~~~0.8 & ~~0.4& 32 \\
16.0 & ~~~~2.3 & ~~0.8& 11\\
16.5 & ~~~~3.3 & ~~0.8& 7\\
17.0 & ~~~~9.5 & ~~1.6& 19\\
17.5 & ~~13.6 & ~~1.9& 13\\
18.0 & ~~26.2 & ~~2.6& 6\\
18.5 & ~~49.9 & ~~3.6& 8\\
19.0 & ~~83.6 & ~~5.6& 7\\
19.5 & 171.5 & ~~8.0& 8\\
20.0 & 315.4 & 13.5& 7\\
20.5 & 596.0 & 18.7& 7\\
21.0 & 931.1 & 23.3& 5\\
\hline
\end{tabular}
\end{center}
\begin{flushleft}
\vspace{2mm}
\hspace{-5mm}{\em a}: N is in unit of $mag^{-1}$ $d^{-2}$.\\
\vspace{2mm}
\hspace{-5mm}{\em b}: $\sigma$ is calculated by using Eq. (5) and (6).\\
\vspace{2mm}
\hspace{-5mm}{\em c}: $\sigma_{ff}$ is the ratio of the field-to-field
variation to the number counts.\\
\end{flushleft}
\newpage
\begin{center}
{\bf Table 4: Slope of No-Evolution Models}\\
\begin{tabular}{c|c|c}
\hline
\hline
$M_*$ & -23.59 & -22.75\\
\hline
kc$^a$ & 0.602 & 0.605\\
kc$^b$ & 0.617 & 0.623\\
kc$^c$ & 0.611 & 0.618\\
\hline
\end{tabular}
\end{center}
\begin{flushleft}
\vspace{2mm}
\hspace{-5mm}{\em a}: the k-correction of Bruzual \& Charlot (1993)\\
\vspace{2mm}
\hspace{-5mm}{\em b}: the k-correction of Rocca-Volmerange \& Guiderdoni 
(1988)\\
\vspace{2mm}
\hspace{-5mm}{\em c}: the k-correction of Cowie {\sl et. al.} (1994)\\
\end{flushleft}
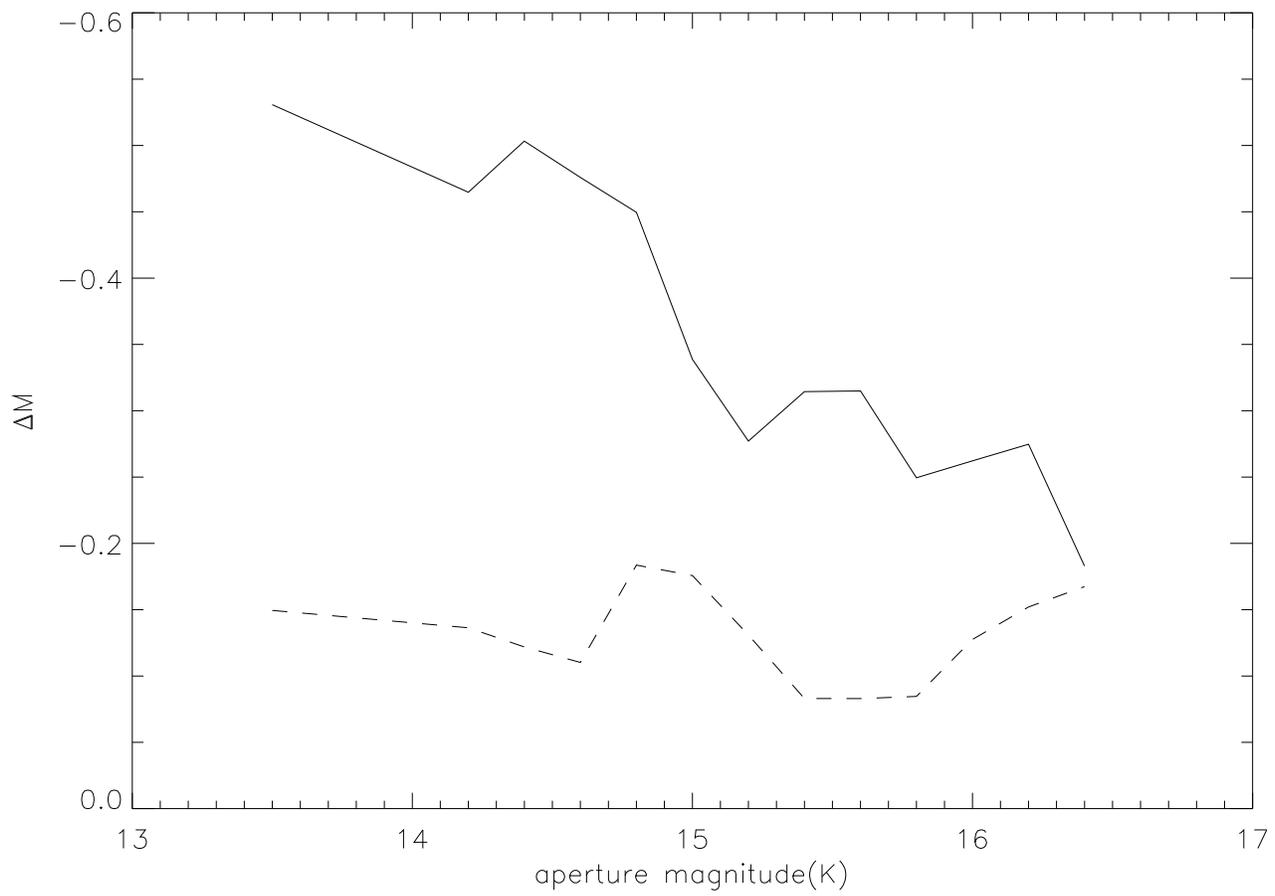
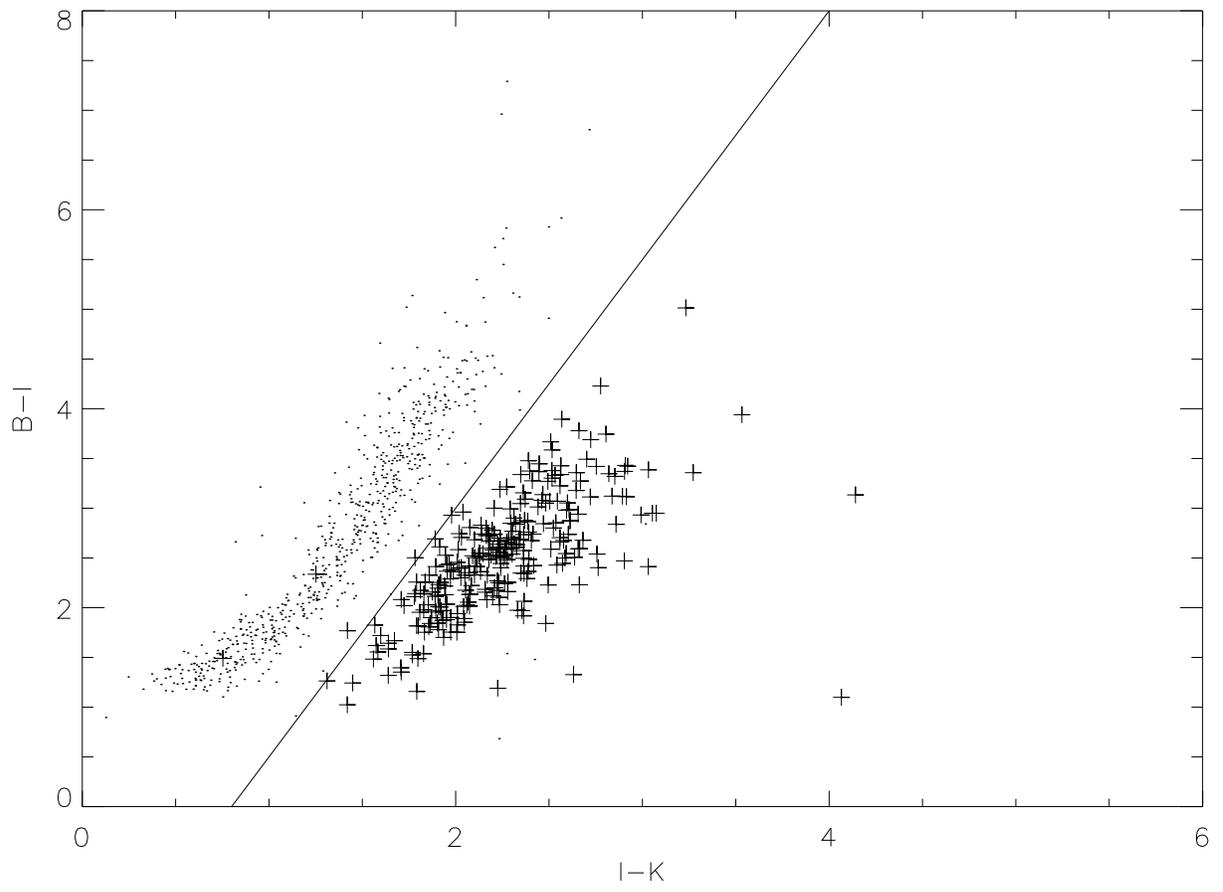
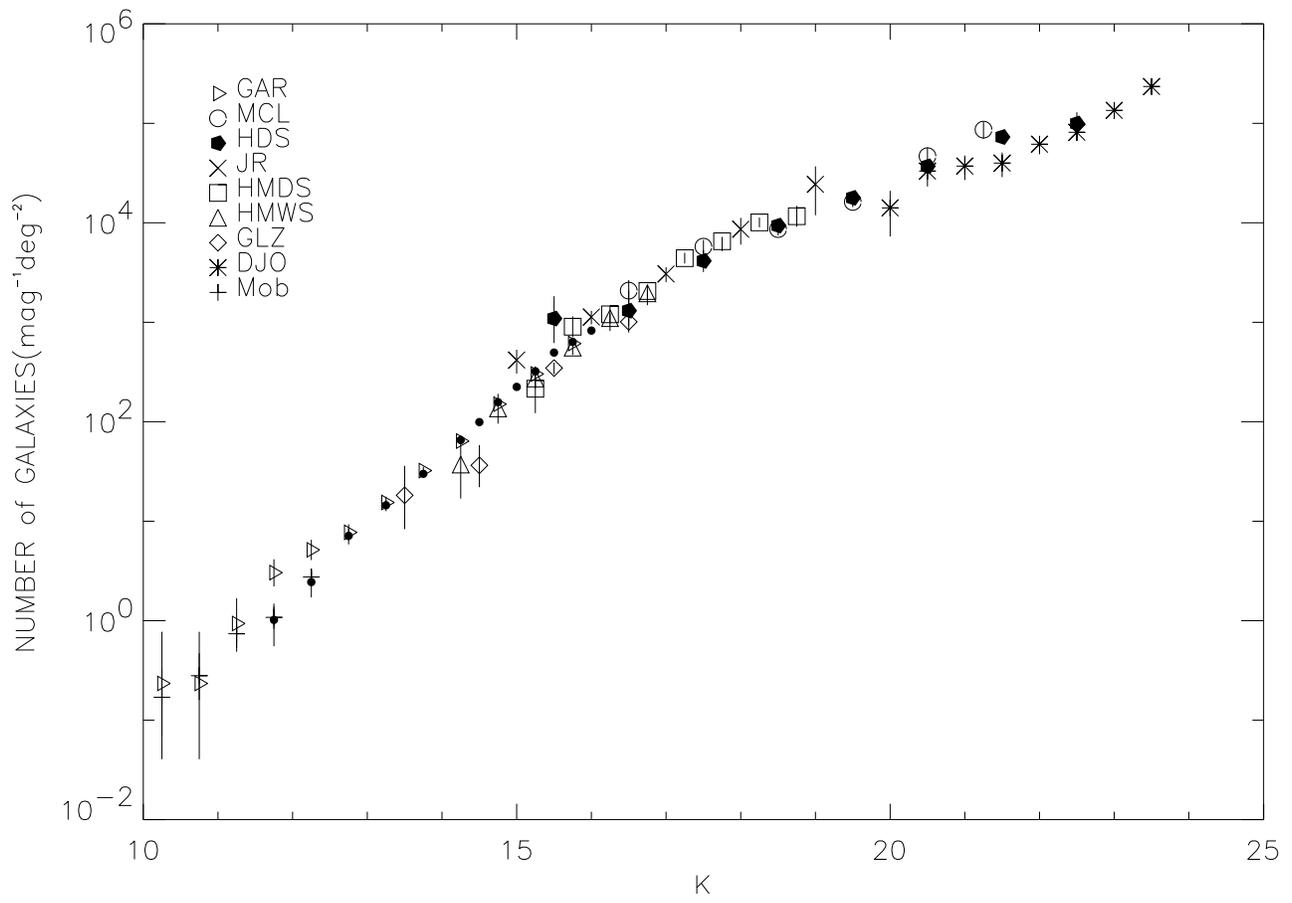
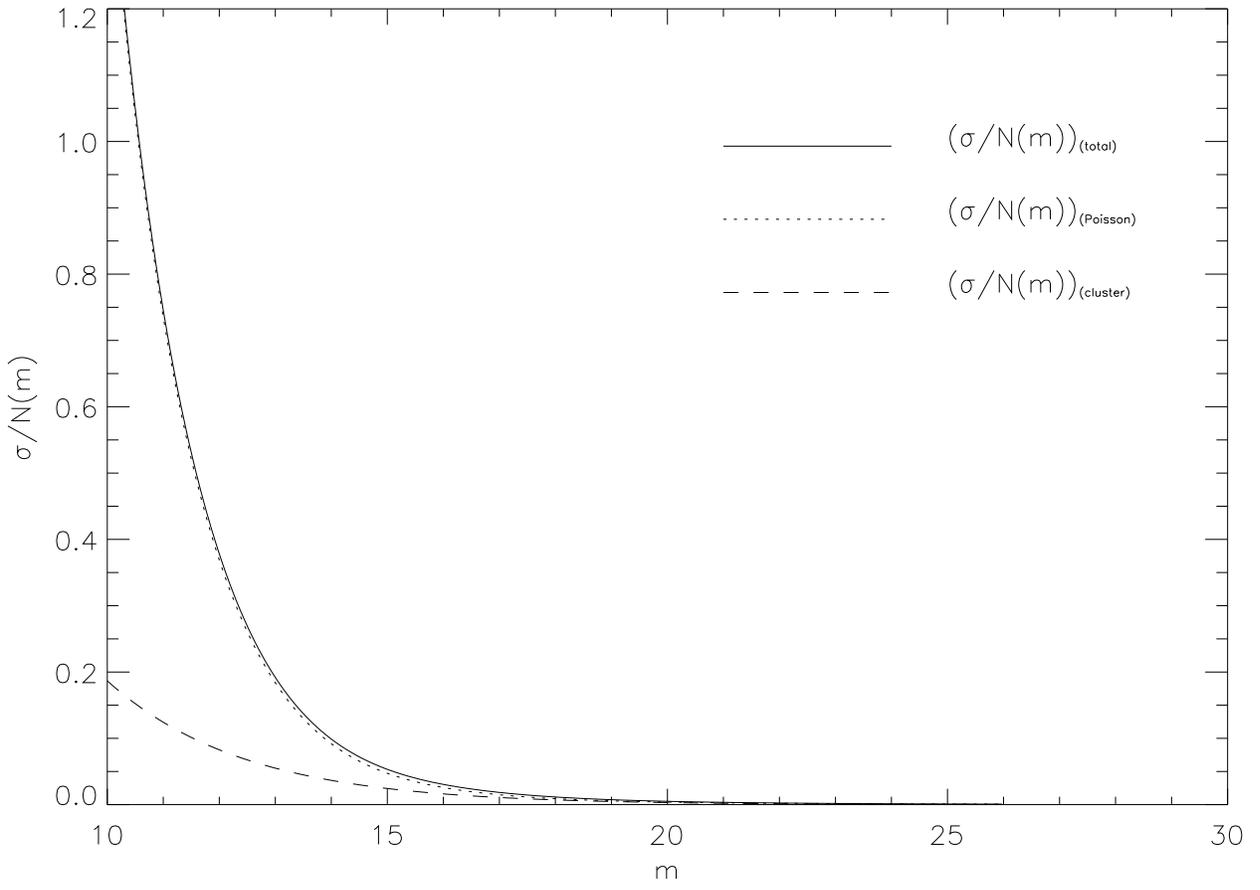
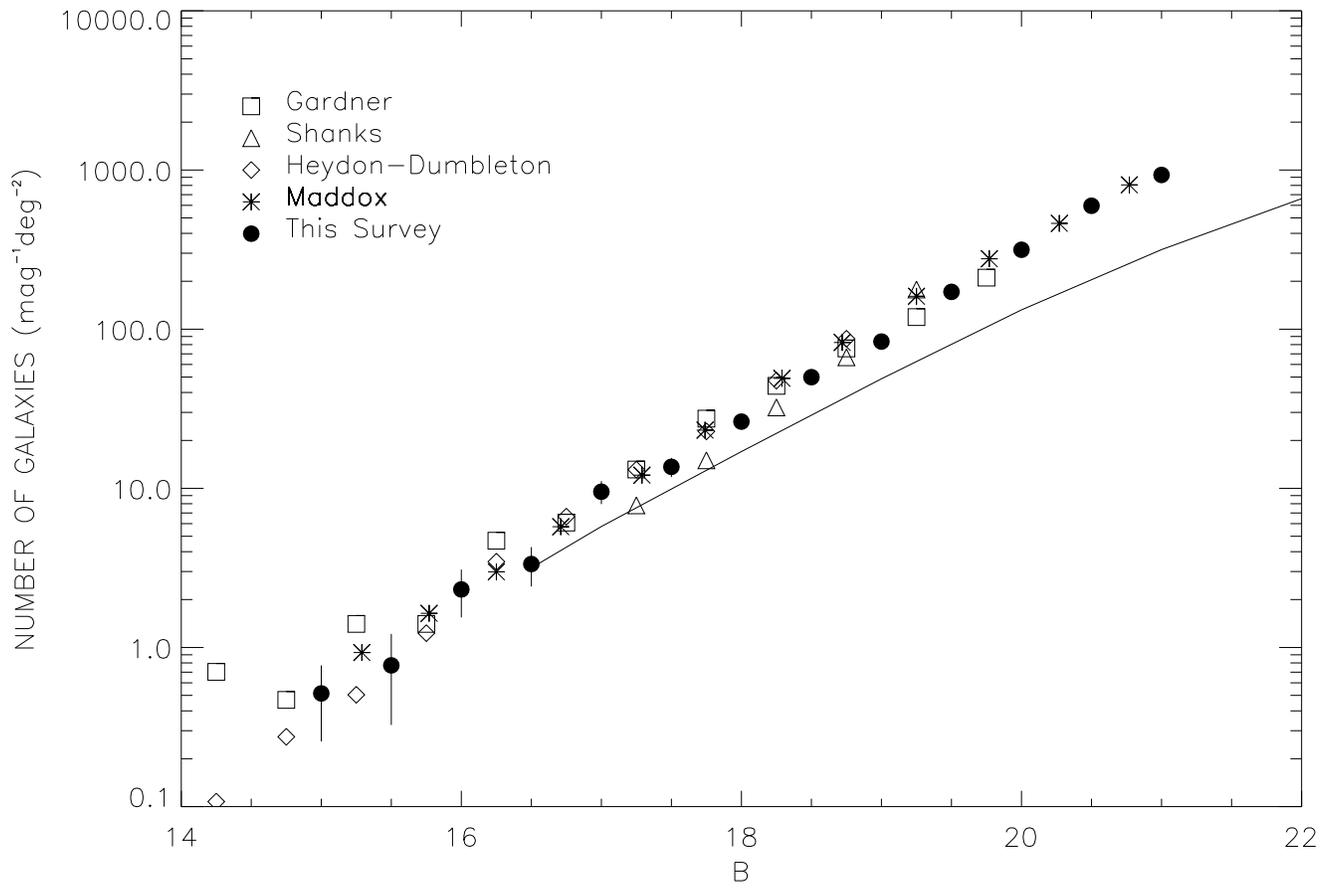
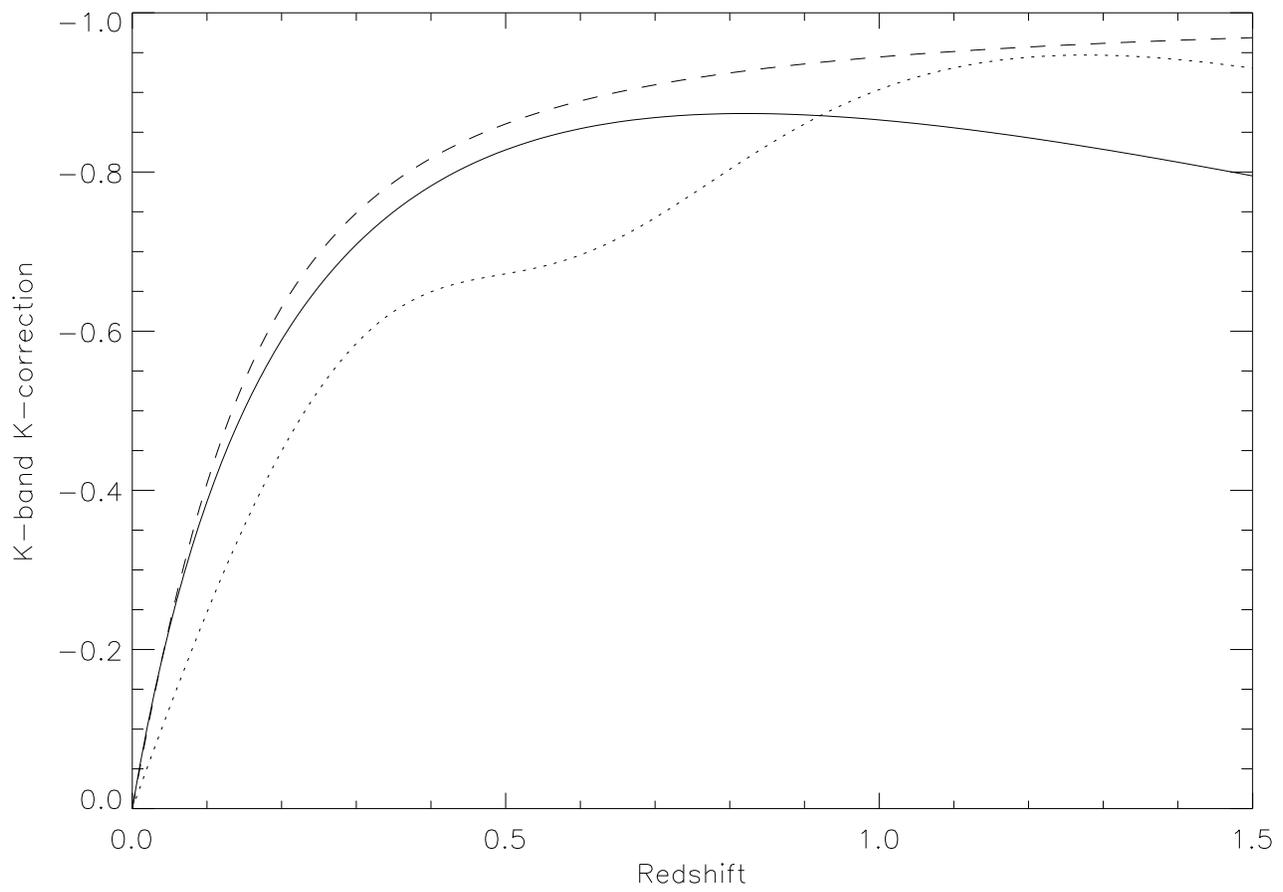
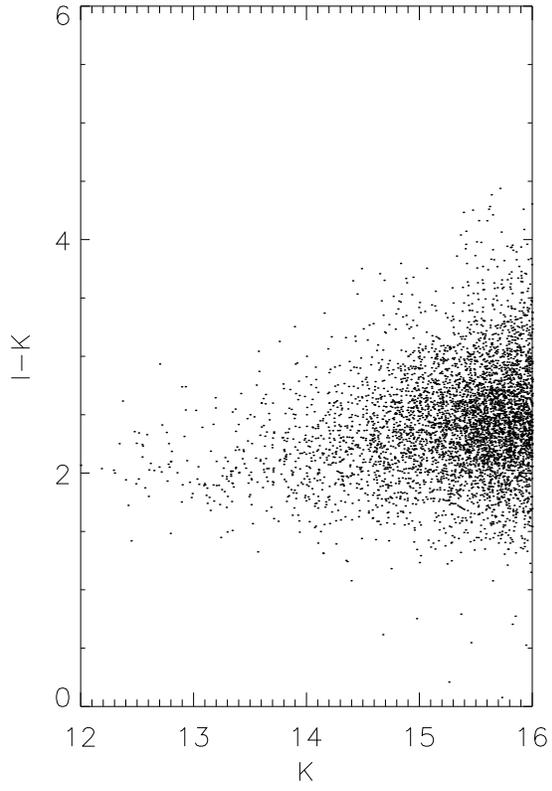
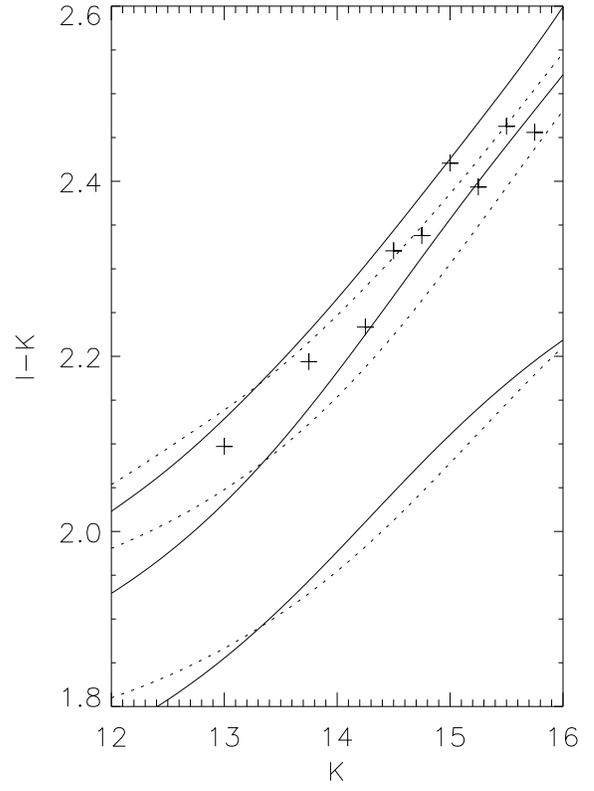
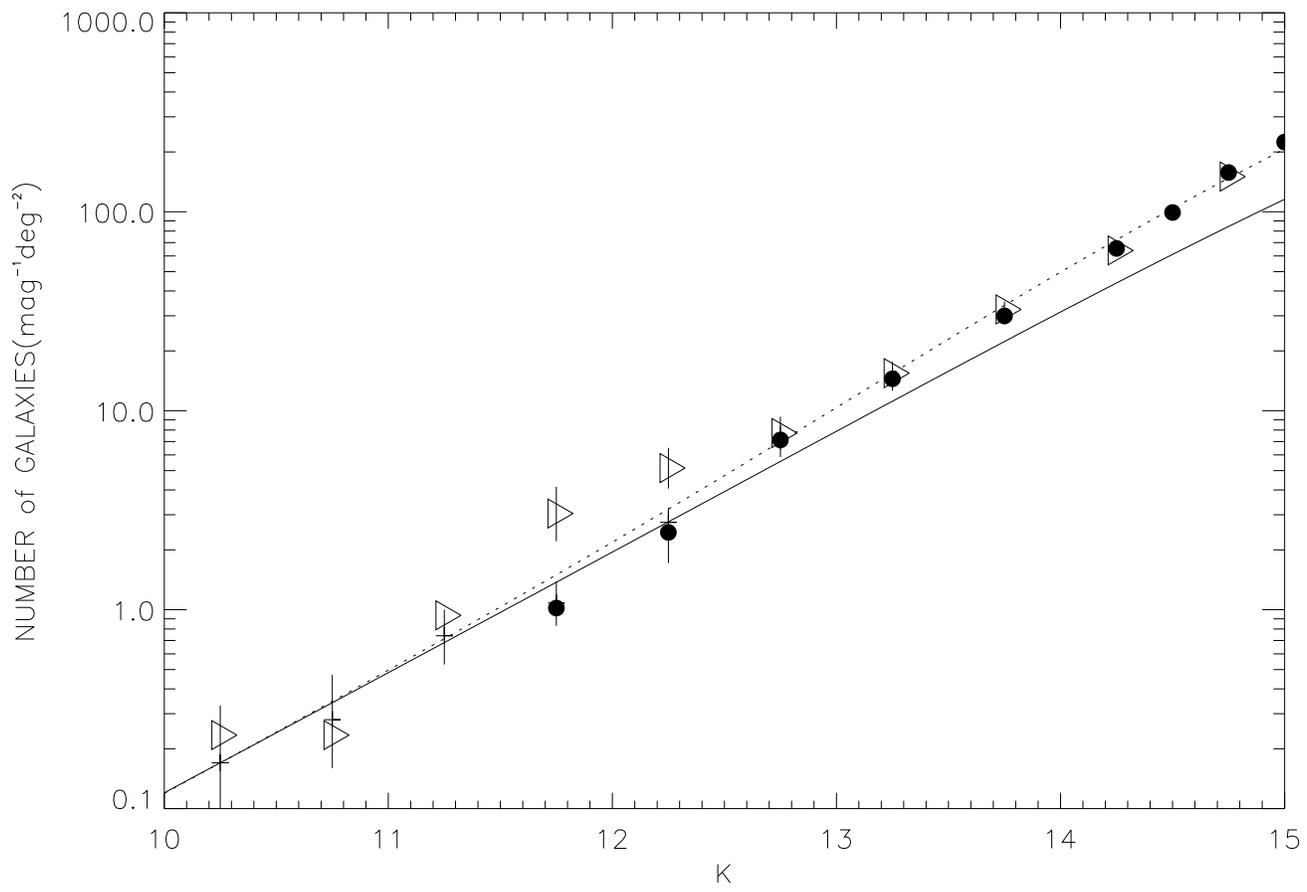
\end{document}